\documentclass[conference]{IEEEtran}

\usepackage{epsfig, amsmath, amssymb}
\begin{document}

\title{A Multitaper, Causal Decomposition for Stochastic, Multivariate Time Series: Application to High-Frequency Calcium Imaging Data}

\author{
\IEEEauthorblockN{Andrew T. Sornborger}
\IEEEauthorblockA{\small Department of Mathematics, University of California, Davis, CA\\
Email: ats@math.ucdavis.edu}
\and
\IEEEauthorblockN{James D. Lauderdale}
\IEEEauthorblockA{\small Department of Cellular Biology, University of Georgia, Athens, GA\\
Email: jdlauder@uga.edu}}

\maketitle

\begin{abstract}
\noindent
Neural data analysis has increasingly incorporated causal information to study circuit connectivity. Dimensional reduction forms the basis of most analyses of large multivariate time series. Here, we present a new, multitaper-based decomposition for stochastic, multivariate time series that acts on the covariance of the time series at all lags, $C(\tau)$, as opposed to standard methods that decompose the time series, $\mathbf{X}(t)$, using only information at zero-lag. In both simulated and neural imaging examples, we demonstrate that methods that neglect the full causal structure may be discarding important dynamical information in a time series.
\end{abstract}


\section{Introduction}
\linespread{1.4}
\IEEEPARstart{T}he technology for imaging neural systems has improved to the point where we can rapidly be overcome by the sheer size of our datasets. For this reason, a significant amount of effort has been expended to develop dimensional reduction methods. Independent component analysis \cite{pmid9673671,pmid10946390}, non-negative matrix factorization \cite{pmid21373259}, generalized linear models \cite{pmid10600421}, principal component analysis \cite{pmid9653187}, wavelet analyses \cite{pmid25629800,pmid20832427}, and other matrix factorizations \cite{pmid18758544} have all been used to advantage in the analysis of neural imaging data.

The lagged-covariance
\begin{equation}
   C(\tau) = E \left\{ \int dt \mathbf{X}(t ) \mathbf{X}^T(t- \tau) \right\}
\end{equation}
and its Fourier transform, the cross-spectrum, 
\begin{equation}
  S(f) = \int C(\tau) e^{-2\pi i f\tau} d\tau
\end{equation}
are fundamental to the description of a stationary random process, $\mathbf{X}(t)$. For a stationary process, the cross-spectrum is uncorrelated as a function of frequency, $f$. Therefore, statistical estimates of stationary processes are often performed in the frequency domain. Since estimates of the cross-spectrum are approximately independent and identically distributed (i.i.d.), confidence intervals can typically be calculated more easily in the frequency domain.

Despite the fact that the Fourier transform can decorrelate a stationary process, uncertainty principles are also fundamental to the analysis of time series. They are a statement that a function that has well-localized support in one set of coordinates, say time, becomes delocalized when transformed to another set of coordinates, i.e. frequency. Therefore, although the statistics of a stationary process are more tractable in frequency coordinates, if the covariance of the process, $C(\tau)$, has more localized temporal support than the cross-spectrum, $S(f)$, then frequency-by-frequency estimation of covariance matrices can result in a loss of statistical power, since the power can be spread across many frequencies. This statement also holds in the opposite direction if the cross-spectrum has more localized frequency support than the covariance.

The goal of this paper is to improve the detection and estimation of lagged-covariance or cross-spectral information using multivariate, multitaper methods \cite{Thomson1982,pmid21970814}. We show that these methods can capture information in the data that goes unseen with standard dimensional reduction methods based only on zero-lag information.

\section{Methods}\label{methods}
To attack this problem, we set up a framework for the detection and estimation of statistically significant covariance or cross-spectral information in multivariate data. The dimensional reduction of covariance or cross-spectral tensors is a technique that we have found to be very useful in revealing the structure latent in time series data. We use multitaper methods to construct our estimates. Multitaper methods provide a set of approximately independent estimates of the covariance or cross-spectrum allowing us to form consistent estimates with low broad-band frequency bias.

\subsection{The Probabilistic Model}
We will assume that the time series of interest may be represented, at least approximately or over a given time window, as a band-limited, weakly stationary process
\begin{equation}
   \mathbf{X}(t) = \int_{-f_N}^{f_N} d\mathbf{Z}(f) e^{2\pi i ft} \;.
\end{equation}
Here, $d\mathbf{Z}(f)$ represents a vector of length $P$ of orthogonal increment Cram\'er processes (c.f. \cite{Thomson1982,Brillinger,Walden2000,Percival_Walden}) for which
\begin{equation}
  E\{ d\mathbf{Z}(f) \} = \mathbf{0}
\end{equation}
and
\begin{equation}
  E\{ d\mathbf{Z}(f)d\mathbf{Z}(f')^\dagger \} = \left\lbrace \begin{array}{ll} S(f)df, & f = f' \\ 0, & f \neq f' \end{array} \right . \; .
\end{equation}
With these assumptions, the covariance at lag $\tau$ may be written
\begin{equation}
  C(\tau) = E \left\{ \int dt \; \mathbf{X}(t) \mathbf{X}^T(t-\tau) \right \} \nonumber
\end{equation}
\begin{equation}
  = E \left\{ \int dt \int_{-f_N}^{f_N} d \mathbf{Z}(f) \int_{-f_N}^{f_N} d \mathbf{Z}^\dagger(f') e^{2 \pi i (f-f')t} e^{2\pi i f\tau} \right\} \; ,
\end{equation}
giving
\begin{equation}
  C(\tau) = \int_{-f_N}^{f_N} df S(f) e^{2\pi i f\tau} \; .  \label{covspec}
\end{equation}
Here and below, bolded letters denote vectors; capital, unbolded letters denote matrices; small, unbolded letters denote scalars; $^T$ denotes the matrix transpose, $^\dagger$ denotes the Hermitian transpose (complex conjugated and transposed) and $E\{ \}$ denotes an expectation value.

\subsection{Causal Decomposition: Decomposing the Lagged-Covariance and Cross-Spectral Tensors}
The most commonly used method for decomposing a data matrix is the singular value decomposition (SVD). This decomposition is used to estimate the principal components of stochastic processes and is so useful in the study of system dynamics (among other things) that it was reinvented many times during the 20th century and is also called the proper orthogonal decomposition (POD), the Karhunen-Lo\`eve (KL) transform and the method of empirical eigenfunctions.

The SVD of a vector time series is a decomposition of the form
\begin{equation}
  \mathbf{X}(t) = \sum_{n=1}^{\mathrm{rank}\mathbf{X}} \mathbf{u}_n d_n v_n(t) \; ,
\end{equation}
where
\begin{equation}
  \int dt \mathbf{X}(t) \mathbf{X}(t)^T \boldsymbol{u}_n = \lambda_n \boldsymbol u_n \; , \label{SVDEig}
\end{equation}
the $\{ \boldsymbol{u}_i \}$ are mutually orthogonal, the $\{ d_i = \sqrt{\lambda_i} \}$ are listed in descending order and $\{ v_i(t) \}$ are the normalized projections $v_i(t) = \boldsymbol{u}_i^T \mathbf{X}(t)/d_i$, and are mutually orthogonal. Note that the operator $\int dt \mathbf{X}(t) \mathbf{X}(t)^T$ on the left of (\ref{SVDEig}) is just the zero-lag covariance matrix, $C(0)$. Other decompositions, such as those mentioned in the Introduction are also based on reductions that use zero-lag information.

Typically, the SVD is first used to find left, $\mathbf{u}_n$, and right, $v_n(t)$, eigenvector bases for $\mathbf{X}(t)$. Then, for dimensional reduction, the standard procedure is to truncate the decomposition by setting a set of small singular values, $\{ d_i \}$, to zero and reconstructing $\mathbf{X}(t)$ in the subspace of the remaining left and right eigenvectors. 

As we saw in the Introduction and above, for stationary processes, the fundamental quantity of interest for modeling a stochastic system is the set of lagged covariance matrices, $C(\tau)$, or the cross-spectral matrices at different frequencies, $S(f)$. These translationally-invariant quantities measure how the stochastic system evolves over time depending on its previous history. Therefore, instead of applying the SVD to the raw data, $\mathbf{X}(t)$, we apply it to estimates of the tensor $C(\tau)$ or $S(f)$ (note that once $\tau$ or $f$ is discretized, these quantities are indeed rank-three tensors). The SVD is not unique for tensors of rank higher than two. However, since $\tau$ and $f$ are clearly special since they are ordered, we will decompose $C(\tau)$ or $S(f)$ along these indices. Thus, the decomposition of the covariance becomes
\begin{equation}
  C(\tau) = \sum_n K_n a_n p_n(\tau) \; , \label{CD1}
\end{equation}
where the $\{ K_n \}$ are eigen-covariance matrices, the $\{ a_n \}$ are singular values and the $\{ p_n(\tau) \}$ are time- or lag-like eigenvectors. Here, $n = 1,\dots,N$, where $N = \mathrm{rank}(\mathbf{X})^2$. An analogous treatment results in the decomposition of the cross-spectrum
\begin{equation}
   S(f) = \sum_n H_n b_n q_n(f) \; , \label{CD2}
\end{equation}
where the $\{ H_n \}$ are eigen-cross spectral matrices, the $\{ b_n \}$ are singular values and the $\{ q_n(f) \}$ are frequency-like eigenvectors.

Since these are decompositions of quantities that are fundamentally related to the causal structure of stationary processes, we will refer to Eqs. (\ref{CD1}) and (\ref{CD2}) as Causal Decompositions (CDs).

\begin{figure}
\centering
\epsfig{figure=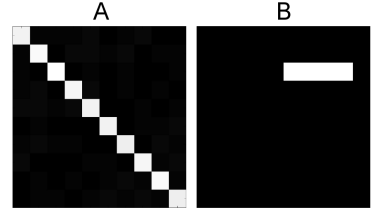,height=1.2in,width=2.0in}
\caption{{The two covariance matrices generated by autoregressive process (\ref{ARDef}). A) $C(0)$, the covariance matrix with support at zero lag; B) $C(20)$, the covariance matrix at lag $\tau = 20$. This matrix is identical to that at $\tau = 40$ and $\tau = 60$.}
\label{ARProcess}}
\end{figure}

\subsection{Multitaper Estimation}
To obtain consistent statistical estimates of $C(\tau)$ and $S(f)$, we resort to multitaper spectral analysis \cite{Thomson1982}. Multitaper methods have been used successfully in many applications in the analysis of neural time series \cite{ObservedBrain}. Multitaper analysis is based on the projection of a time series onto a set of $m = 1, ..., M$ orthogonal `tapers' called Slepians (also called discrete prolate spheroidal sequences or DPSSs). Slepians are a complete set of functions that are ordered in terms of their frequency concentration and serve as optimal band-pass filters provided the maximum index $M < 2TW-3$. Here, $T$ is the number of points in the time series and $W$ is a user-defined frequency bandwidth. The data is first tapered, then Fourier transformed, resulting in a set of $M$ frequency estimates, so-called eigenestimates, with low broad-band bias (i.e. estimates of amplitudes at a given frequency are relatively uninfluenced by frequencies outside of the bandwidth $W$).

The data is averaged across the eigenestimates, resulting in smoothed spectral estimates with reduced variance (relative to a single, tapered periodogram). 

Because multitaper techniques rely on a pre-defined bandwidth $W$, the CDs described here may be smoothed by the user to different degrees. More smoothing gives rise to smaller confidence intervals, but estimates that have lower resolution in frequency, whereas less smoothing results in larger confidence intervals, but higher frequency resolution.


Define the tapered eigenestimate
\begin{equation}
  \mathbf{J}_m(f) \equiv \sum_{t=1}^T h_m(t) \mathbf{X}(t) e^{-2\pi i ft} \; ,
\end{equation}
where $h_m(t)$ is a Slepian function. A multitaper estimate of $S(f)$ is given by
\begin{equation}
  \hat{S}(f) = \frac{1}{M-1} \sum_{m = 1}^{M} \mathbf{J}_m(f) \mathbf{J}_m(f)^\dagger
\end{equation}

$\hat{C}(\tau)$ may then be estimated via (\ref{covspec}).

\begin{figure}
\centering
\epsfig{figure=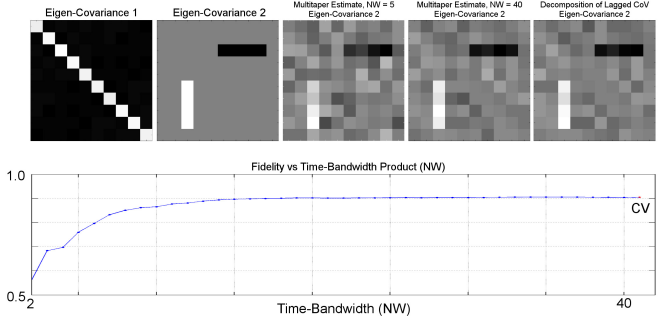,height=1.75in,width=3.5in}
\caption{
{A) The exact simulated AR process covariance at zero-lag; B) The exact AR process eigen-covariance at lags $20$, $40$, and $60$. Note that this is equal to $A - A^T$ (see text); C,D) The second estimated eigen-covariances from CDs computed using multitaper methods in the frequency domain with time-bandwidth products TW = 5 and 40, respectively. Note that the first eigen-covariance was to very close approximation the identity matrix; E) Estimated second eigen-covariance from a non-multitaper CD computed in the time (lag) domain; F) Fidelity (correlation of estimated with exact eigen-covariance) of second eigen-covariance as a function of TW. `CV' denotes the fidelity of the non-multitaper estimate taken in the time (lag) domain. Here, we see that $TW \sim 15$ is sufficient for good esimation. For these simulations, $T = 10,000$.}
\label{ARProcess}}
\end{figure}

\subsection{Statistical Significance}
The estimate $\hat{S}(f)$ is Wishart distributed with $2(M-P)$ degrees of freedom \cite{pmid21970814}. However, the distribution of $\hat{C}(\tau)$ is more complicated, due to the Fourier transform in (\ref{covspec}). Furthermore, we wish to obtain confidence intervals on the eigen-cross spectral matrices, singular values and frequency-like eigenvectors of the SVD of $S(f)$ (and the corresponding objects for the decomposition of $C(\tau)$). Jackknife (leave-one-out) estimates are well-suited for application in these circumstances. First, we outline the jackknife estimation procedure for $S(f)$. We define
\begin{equation}
  \hat{S}^j(f) = \frac{1}{M-1} \sum_{(j)} \mathbf{J}_m(f) \mathbf{J}_m(f)^\dagger \; .
\end{equation}
where $\sum_{(j)}$ indicates the sum over all $M$ tapers except the $j$'th taper. We now perform an SVD on $\hat{S}^j(f)$ resulting in sets of eigen-covariance matrices, $\{ \{ H_i \}^j \}$, singular values, $\{ \{ b_i \}^j \}$, and frequency-like eigenvectors, $\{ \{ q_i \}^j \}$, where $i = 1,\dots,N$ and $j = 1,\dots,M$.

From these sets, we calculate estimated means and variances for $\{ H_i \}$, $\{ b_i \}$, and $\{ q_i \}$. With estimates of the mean and variance, it is straightforward to calculate confidence intervals for the eigen-covariance matrices, eigenvectors and singular values.


This procedure for calculating mean and variance estimates of the objects resulting from the decomposition of $S(f)$ may also be used to calculate mean and variance estimates of the objects resulting from the decomposition of $C(\tau)$. In this case, we perform SVDs on $C^j(\tau)$, where
\begin{equation}
   \hat{C}^j(\tau) = \int df \hat{S}^j(f) e^{2\pi i f\tau} \; ,
\end{equation}
and calculate the jackknife estimates as we did for $S^j(f)$. Notice that even though $\hat{C}^j(\tau_1)$ is not independent of $\hat{C}^j(\tau_2)$, the sample $\hat{C}^{j_1}(\tau)$ is independent of $\hat{C}^{j_2}(\tau)$ because it is calculated with Slepian tapers whose Fourier transform is orthogonal to the other Slepian tapers in the frequency domain. Thus the samples are approximately i.i.d.. This is an example of a transformation-based-bootstrap method \cite{TransformationBasedBootstrap}, which leaves out independent samples in the frequency domain.

For practical purposes, the CDs in (\ref{CD1}) and (\ref{CD2}) may be computed as `raw' estimates, i.e. without an attempt to estimate statistics of the process. The use of multitaper methods does increase the time required to calculate a CD. Therefore, for cursory or preliminary analyses, it should be remembered that the number of tapers may be decreased or multitaper methods can be done away with entirely. The CD may be computed by simply forming a `raw' estimate of $C(\tau)$ or $S(f)$, then performing the SVD on the raw estimate. In this case, we have no confidence intervals, but we can get a sense of which eigen-matrices, $\{ H_n \}$ or $\{ K_n \}$, represent the dominant contributions to the covariance or cross-spectrum, the form of the singular value spectrum, and the form of the eigenvectors $\{ p_n(\tau) \}$ or $\{ q_n(f) \}$.
 
\section{Results}

We tested our method on both simulated and actual neural imaging data. The simulated data analysis is meant to demonstrate a situation in which a standard data reduction would not detect any statistically significant structure, but a CD should find statistically significant correlations in the data. The neural imaging data that we analyze are 1000Hz, fluorescence line-scan measurements of seizure-related calcium activity using the ratiometric, genetically-encoded calcium indicator {\it cameleon} YC2.1 in a larval zebrafish. We demonstrate that with this data set our method detects structure that would not have been detected with a standard analysis.

\subsection{Simulated Data}

We used an autoregressive (AR) process to generate a stationary multivariate time series, $\mathbf{X}(t)$. The process was
\begin{equation}\label{ARDef}
   \mathbf{X}(t) = A \mathbf{X}(t - 20) + A \mathbf{X}(t - 40) + A \mathbf{X}(t - 60) + \mathbf{N}(t) \; ,
\end{equation}
where the matrix $A$ is depicted in Fig. 1B and $\mathbf{N}(t)$ is an i.i.d. Gaussian random vector. Note that with this matrix, there was no feedback in the time series. Time series $6$ through $9$ were simply added to time series $3$ at the fixed lags of $20$, $40$ and $60$ time units. Therefore the covariance tensor of this time series was diagonal at $\tau = 0$ (Fig. 1A) (capturing the variance of each time series), proportional to $A$ at lags $20$, $40$, and $60$ and proportional to $A^T$ at lags $-20$, $-40$, and $-60$. Because all variables in the process have the same variance, $C(0)$ is proportional to the identity matrix and an SVD of this data, whose left singular vectors are eigenvectors of $C(0)$, outputs random right/left singular vectors with ordered, normally distributed singular values.

In Fig. 2, we present results from a CD of the simulated AR process. The first eigen-cross-spectral estimates of the multitaper CD captured the diagonal $C(0)$ cross-spectral matrix (see Fig. 1A for the theoretical matrix, the first eigen-cross-spectral matrix estimate is not shown). The second eigen-cross-spectral matrix of the multitaper CD captured the covariance due to A (Fig. 2B, this matrix is $H_2 \sim A - A^T$), with estimation improving as $TW$ increased (Fig. 2C,D). Note also that, since the covariance tensor of this time series only has narrow ($\delta$-function) support at the discrete lags $0$, $\pm 20$, $\pm 40$, $\pm 60$, the uncertainty principle predicts that the cross-spectral tensor will have broad support across many frequencies, but the covariance tensor will have sharp ($\delta$-function) support as a function of time (lag). For this reason, the `raw' CD in the time (lag) domain is more statistically efficient than low $TW$ multitaper estimators (in the sense of smallest mean squared error, Fig. 2F) due to the few, high signal-to-noise peaks in covariance support (Fig. 1E, red point in F). However, we have no confidence intervals for such raw estimators. We would expect processes with narrow frequency support to be more efficiently computed in the frequency domain.

\begin{figure}
\centering
\epsfig{figure=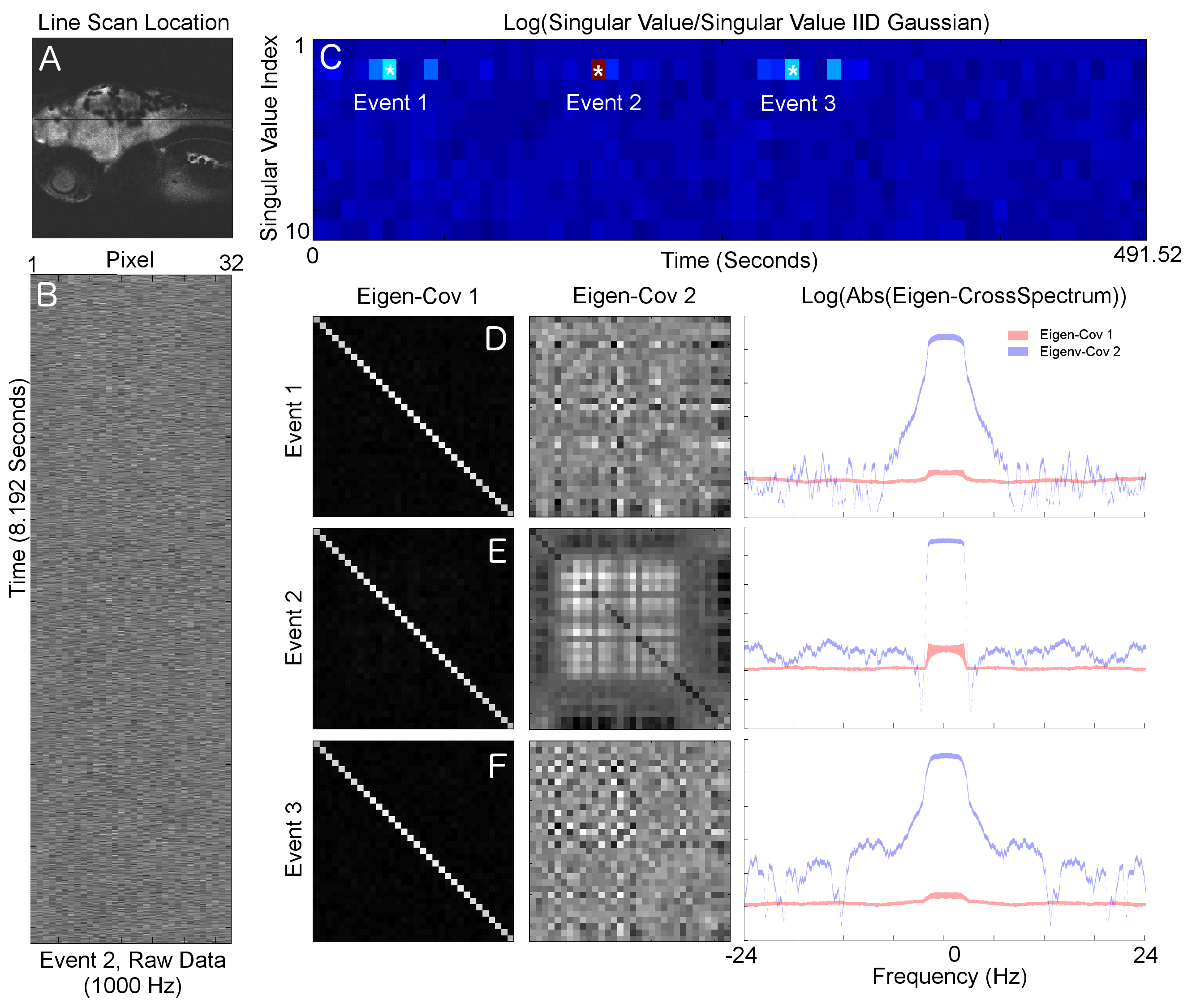,height=3in,width=3.5in}
\caption{A) Location of line-scan in 9 dpf larval zebrafish; B) Raw imaging data segment (one of 60 taken at 1000Hz); C) Normalized singular values from multitaper CD (see text); D, E, F) Eigen-cross-spectral matrices and eigen-cross-spectra from the three events labelled with asterisks in C). The eigen-cross-spectra are plotted with $2$-$\sigma$ confidence intervals.)
\label{HighFrequency}}
\end{figure}

\subsection{High-frequency Calcium Imaging Data}

Calcium imaging data taken at high frequencies are capable of detecting individual action potentials, but at the cost of a reduced signal-to-noise due to the reduced number of photons per image in an acquisition. Thus, dimensional reduction methods are important for the analysis of such data. Action potentials {\it in vivo} are typically modeled as a stochastic point process. Because individual neurons communicate with each other with synaptic propagation times on the order of tens of milliseconds, one expects that a data-reduction using an SVD would not give useful results since only zero-lag correlations are being taken into account. This is indeed the case with the data set that we analyzed here. However, using a multitaper CD, we found a number of statistically significant events that went undetected with the SVD analysis.

The line-scan calcium imaging data that we analyzed was taken from a line through the dorsal central nervous system of a 9 days-post-fertilization (dpf) larval zebrafish (Fig. 3A). The data was taken in a set of 60 contiguous 8.192 second segments (Fig. 3B) for a total of 491.52 seconds ($\sim 8$ minutes). A multitaper CD was performed on each segment. In order to visualize statistically significant events, the spectrum of a data matrix with normally distributed noise of equal power to the data in a given segment was subtracted from the spectrum of the CD (Fig. 3C). A series of statistically significant events were detected, three of which are shown in Fig. 3D-F in more detail. Each of these panels shows the two most significant eigen-cross-spectral matrices and the logarithm of the absolute value of the eigen-cross-spectrum. Note that the first eigen-cross-spectral matrix is, to a very good approximation, the identity. This is the matrix that would be diagonalized by the right eigenvectors of an SVD and would fail to identify the cross-spectral information contained in the second (and subsequent) eigen-cross-spectral matrices and eigen-cross-spectra.

The eigen-matrices and eigen-spectra show different structure from event to event. We will not interpret this here because it is impossible to interpret causal relationships in line scan data; the whole neuronal circuit is not sampled. However, due to the variety of relationships between line-scan pixels in the data, it is clear that a rich variety of neural mechanisms could be at play. Given a complete sample of neurons, using, for instance light-sheet microscopy, one would be able to make better guesses at the neural circuits that underlie these structures.

\section{Discussion and Conclusions}
\enlargethispage{-0.5in}

We have shown that the causal decomposition presented here can detect information that would normally go unnoticed in the standard multivariate dimensional reduction methods that are used to analyze imaging data.

The CDs described here are a useful way of summarizing the information contained in the covariance and cross-spectrum. Like the standard SVD, which may be truncated to denoise a dataset $\mathbf{X}(t)$, the CD may be used to denoise $C(\tau)$ or $S(f)$. The results of a CD may be readily visualized to give the user a clearer understanding of the underlying covariance or cross-spectral structure of a time series. They may also be used to obtain improved estimates of multivariate auto-regressive processes and, hence, improved predictions from measurements of the process. If the user wants to use the CD to dimensionally reduce the dataset (not the covariance or cross-spectrum), since the eigen-cross-spectra are hermitian, $S(f) = S^\dagger(-f)$, $S(f) + S^\dagger(-f)$ (alternatively, $C(\tau) = C(-\tau)^T$, thus $C(\tau) + C(-\tau)^T$) will have an orthogonal, set of real eigenvectors. These may be listed in order of their covariance, thresholded, and the dataset may be projected on them, giving a reconstruction of the data retaining causal information.

We have made no attempt to optimize the decompositions. Improvements could be made for instance because the matrices $S(f)$ are hermitian and $S(f) = S^\dagger(-f)$. Similarly, $C(\tau) = C^T(-\tau)$. We note that a CD may be also be useful when performed on the coherency, amplitude spectrum, or other objects often investigated in the analysis of time series, instead of the cross-spectrum.

By using multitaper techniques, our approach gives us statistics that may be used to obtain confidence intervals on the various objects (eigen-covariance, singular values and lag- or frequency-like eigenvectors) that result from the decomposition. They may also be used to test whether the objects resulting from the decomposition, for instance the singular value spectrum, of a given covariance or cross-spectrum are statistically different from others. As an example, one can form a two-sample $t$-statistic to test for non-stationarity of a process by calculating CDs of two different time-windows of the process and testing whether the distribution of singular values differs between them.

\section*{Acknowledgment}
This work was funded by the National Institutes of Health, CRCNS program NS090645 (ATS and JDL).

\bibliographystyle{IEEEtran}
\bibliography{biblio}

\newpage

\end{document}